\RequirePackage{fix-cm}
\documentclass[twocolumn]{svjour3}          
\usepackage{color}
\usepackage[margin=2cm]{geometry}
\usepackage{graphicx}
\usepackage{nameref}
\usepackage{amsmath}
\usepackage{hyperref}
\usepackage{varioref}
\usepackage{booktabs}  
\usepackage[round, sort, comma]{natbib}
\usepackage{mathptmx}      
\usepackage{latexsym}
\begin{document}

\title{Harmonic Sum-based Method for Heart Rate Estimation using PPG Signals Affected with Motion Artifacts}
%
%
\author{Harishchandra Dubey\thanks{\textcolor{blue}{This material is presented to ensure timely dissemination of scholarly and technical work. Copyright and all rights therein are retained by the authors or by the respective copyright holders. The original citation of this paper is:
			H. Dubey, R. Kumaresan, K. Mankodiya, "Harmonic Sum-based Method for Heart Rate Estimation using PPG Signals Affected with Motion Artifacts", Journal of Ambient Intelligence and Humanized Computing, Springer,Oct. 2016.}}  \and Ramdas Kumaresan \and Kunal Mankodiya.
	\institute{H. Dubey is with the Center for Robust Speech Systems, The University of Texas at Dallas, 800 West Campbell Road, Richardson, TX-75080, USA. \\
		R. Kumaresan and K. Mankodiya are with the Department of Electrical, Computer and Biomedical Engineering, The University of Rhode Island, Kingston, 4 East Alumni Ave, Kelley Annex - A215, Kingston, RI 02881, USA.}
	\\Tel.: +1-401-874-5661\\
	Fax: +1-401-874-5661\\
	\email{harishchandra.dubey@utdallas.edu, kunalm@uri.edu, kumar@ele.uri.edu}}         
\maketitle
\begin{abstract}
Wearable photoplethysmography (WPPG) has recently become a common technology in heart rate (HR) monitoring. General observation is that the motion artifacts change
the statistics of the acquired PPG signal. Consequently, estimation of HR from such a corrupted PPG signal is challenging. However, if an accelerometer is also used to acquire the acceleration signal simultaneously, it can provide helpful
information that can be used to reduce the motion artifacts in the PPG signal. By dint
of repetitive movements of the subject’s hands while running,
the accelerometer signal is found to be quasi-periodic. Over
short-time intervals, it can be modeled by a finite harmonic
sum (HSUM). Using the harmonic sum (HSUM) model, we obtain an estimate of
the instantaneous fundamental frequency of the accelerometer
signal. Since the PPG signal is a composite of the heart rate
information (that is also quasi-periodic) and the motion artifact,
we fit a joint harmonic sum (HSUM) model to the PPG signal. One of the
harmonic sums corresponds to the heart-beat component in
PPG and the other models the motion artifact. However, the
fundamental frequency of the motion artifact has already been
determined from the accelerometer signal. Subsequently, the HR
is estimated from the joint HSUM model. The mean absolute
error in HR estimates was 0.7359 beats per minute (BPM) with
a standard deviation of 0.8328 BPM for 2015 IEEE Signal Processing (SP) cup data. The ground-truth HR was obtained from the simultaneously acquired ECG for validating
the accuracy of the proposed method. The proposed method is compared with four methods that were recently developed and evaluated on the same dataset.
\end{abstract}
\section{Introduction}
\label{intro}
Use of wearable sensors such as wrist-bands and smartwatches for monitoring vitals stats of the patients and/or healthy individual is a common trend in recent years~\citep{dubey2015multi, dubey2015fog, dubey2015echowear, monteiro2016fit, dubey2016fogcare}. Wearable body sensors are trending for telemonitoring of personalized health parameters such as heart rate (HR), activity, sleep quality, steps, and calories burned. Wearable sensors have been used in education sector such as analysis of peer-led team learning groups~\citep{dubey2016robust,dubey2016speaker}. Even though, the design of personalized wearable devices is becoming more elegant and user-friendly, their performance is questionable with respect to data reliability. For example, Spierer et al.~\citep{r41} reported that commercial wearable sensors under performed significantly in monitoring HR during daily life activities such as walking, biking, and stair climbing. They used commercial, wearable photoplethysmography (WPPG) sensors that are nowadays found in smartwatches and wristbands~\citep{r4}. The WPPG is a non-invasive technology to capture the cardiac rhythm and hence could be  used for continuous HR monitoring. The WPPG technology is an alternative to electrocardiogram (ECG) for continuous real-time HR monitoring~\citep{r4}-\citep{r1}. As reported in~\citep{r41}, WPPG is significantly affected by the body movements during various activities of daily life. In general, the motion artifacts corrupt the signal components of heart rate or PPG signal. Therefore, HR estimation from wearable PPG signal is a challenging problem faced by the researchers in academia and industry. 

In this  article, we demonstrate our approach of using harmonic sum (HSUM) models in reducing the impact of motion artifacts and in extracting HR accurately from a WPPG signal. We outline the basic principle of a WPPG and related works in Section 2. The motivation and proposed method are described in Section 3, followed by a  discussion on evaluation results in Section 4. Our results demonstrate that the HSUM models are suitable for estimating heart rate (HR) from PPG signals that are severely affected with motion artifacts. 
\begin{figure*}[!t]
\centering
\includegraphics[width=0.99\textwidth]{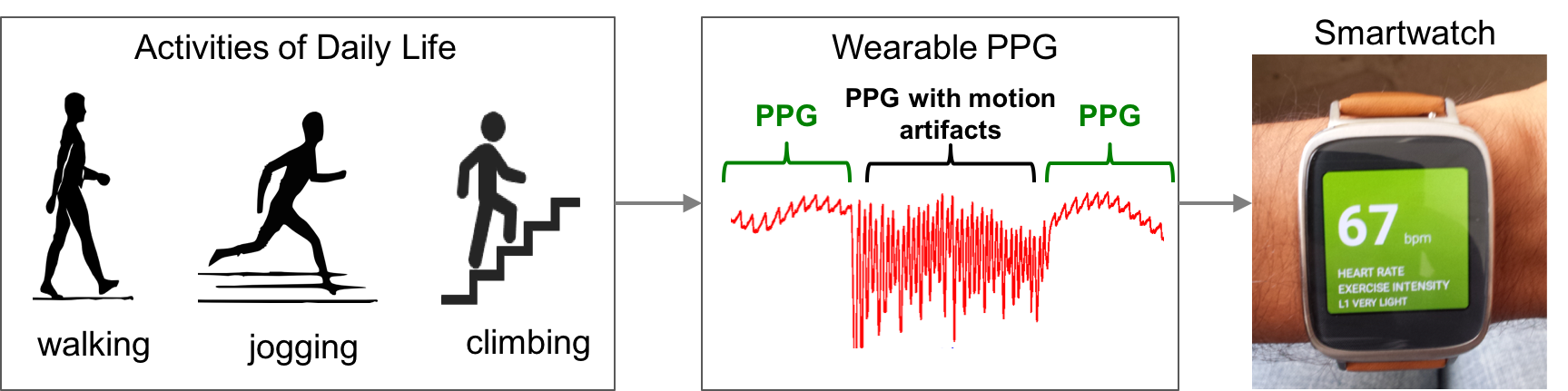}
\caption{Typical scenario for application of commercial wearable PPG sensors.}\label{fig_scenario}%
\end{figure*}
\section{Related Works}
\label{sec:related}
\subsection{Wearable PPG System}
\begin{figure*}[!t]
\centering
\includegraphics[width=0.99\textwidth]{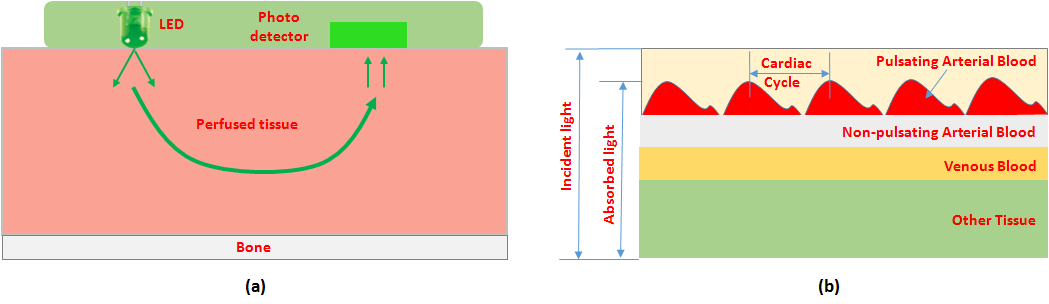}
\caption{(a) Diagram showing the principle of reflectance-type PPG (b) Relation between changes in light intensity and the cardiac cycle (adopted from~\citep{r15}).}
\label{fig_ppg}
\end{figure*}
A wearable PPG system can be either of transmittance-type
or reflectance-type. It consists of a light source
and a detector packed with supporting hardware into a
wristband or earring. The transmittance-type detects the light
transmitted through the tissues by a photodiode kept opposite
to the light source. The reflectance-type detects the
intensity of reflected light using a photodiode kept on the same side as the light source. It works on the change in light intensity upon reflection from the tissue or blood vessels~\citep{r13}. Using PPG sensors on fingertips facilitates a good quality PPG signal with  transmittance-type PPG,
 though it interferes with various activities of daily life. The reflectance-type wristband PPG is often preferred, as it provides least hindrance during different
 activities  and can be incorporated in a smartwatch. Red, infrared or green light-emitting diodes are common light sources in WPPG systems. Figure~\ref{fig_ppg} shows the principle of reflectance-type PPG. The combination of a light source and a light detector is kept together close to the skin surface. The light-emitting diode (source) illuminates the skin that transmits the light. Subsequent tissues partly absorb and partly transmit the light passed through the skin
 surface. Finally, transmitted light is reflected by the subcutaneous tissue. The photodiode (sink) is activated by reflected light generating a voltage signal. The
 voltage signal produced by the photodiode is acquired and filtered by a hardware circuit. The voltage thus acquired is the PPG signal. It is quantitatively related to changes in blood volume in the microvascular tissues. Since it is related to the cardiac rhythm, it can be used for estimation of heart rate (HR). Modern smartwatches are equipped with a  PPG sensor and various other  geometric sensors such as an accelerometer, a gyroscope, and a magnetometer. An accelerometer sensor is commonly found in such smart wristband devices to record the acceleration signal that helps monitor  motion artifacts. The DC component of the  frequency transformed PPG signal corresponds to light absorption from skin, tissues, bones and vascular elements (non-pulsating arterial blood and venous blood). However, the AC component results from pulsating arterial blood flow that is related to cardiac rhythm (systole and diastole). The heart rate can be extracted from the AC component of PPG signal as it is related to the cardiac cycle~\citep{r13}.
\begin{figure*}[!t]
\centering
\includegraphics[width=0.99\textwidth]{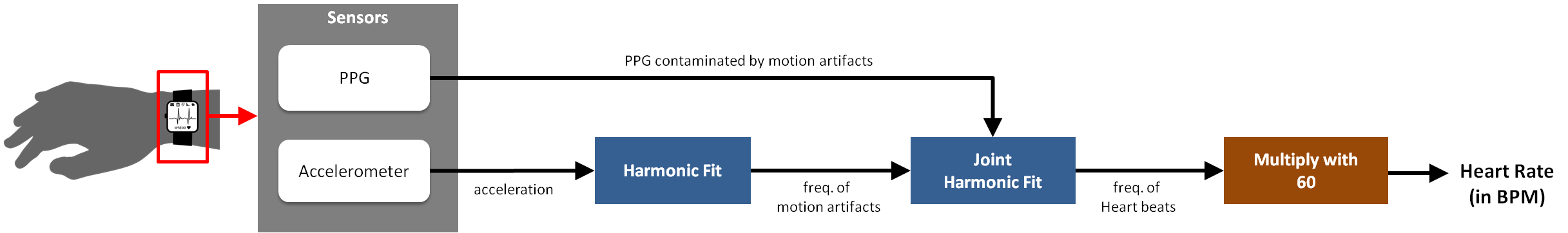}
\caption{The proposed framework for Heart Rate (HR) estimation from the PPG signal affected with strong motion-artifacts. The algorithm take the PPG signal and the accelerometer signal as input and outputs the estimated HRs for each 8-second window with 6-second overlap between successive windows.}
\label{fig_hsum}
\end{figure*}
\begin{figure*}[!t]
\centering
\includegraphics[width=0.99\textwidth]{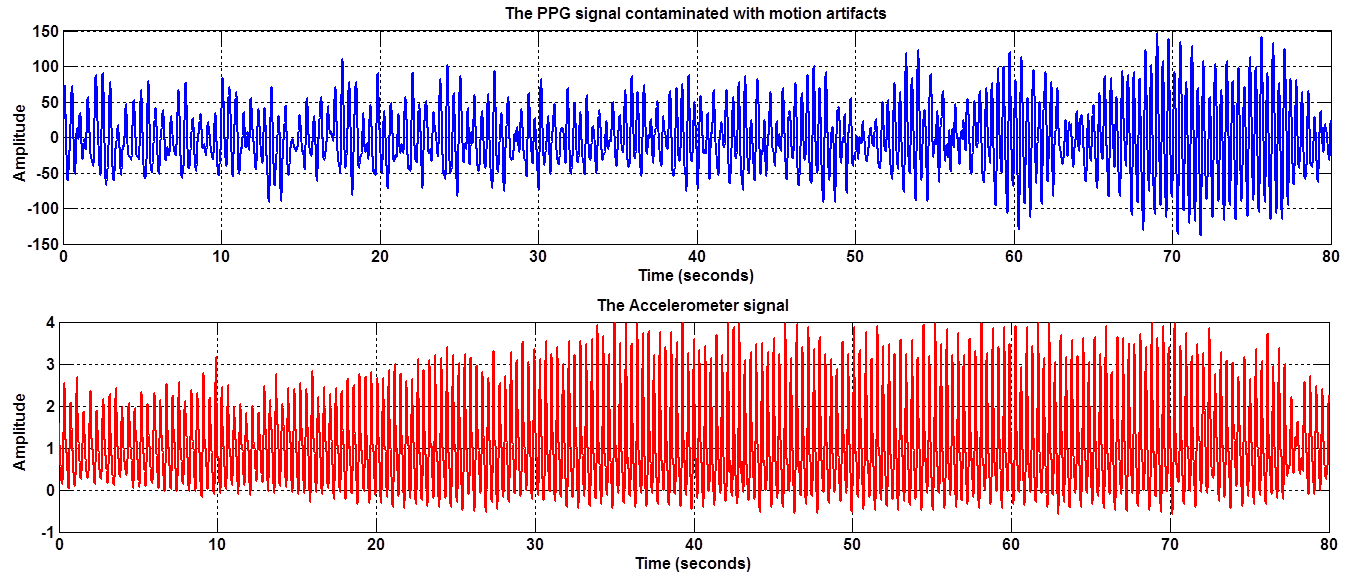}
\caption{An example of the PPG signal corrupted by the motion artifacts (top panel) and  an accelerometer signal collected simultaneously (bottom panel) are shown for an 80-second time-window. The sampling rate is 125 Hz. The quasi-harmonic structure of the accelerometer signal is depicted for this figure. The PPG signal also has a quasi-harmonic structure but an envelope modulation is also observed. Such a  modulation is caused by the interaction between the true heart rhythm signal and the signal components induced by the  physical movements. Hence, the PPG signal may be modeled as a  sum of two harmonic series with slightly different fundamental frequencies over short time-windows. A portion of the DATA05TYPE02 dataset of IEEE SP cup was used for generating this figure~\citep{r10}.}
\label{fig_time}
\end{figure*}%
\subsection{Motion-Artifacts and Distortions in a PPG Signal}
Figure~\ref{fig_scenario}  shows the typical scenario for application of commercial wearable PPG sensors. A PPG signal is corrupted by the influence of external factors
such as ambient light, ambient temperature and pressure in
addition to movements caused by the day to day physical activities. Motion-artifacts (MAs) and pressure
disturbances are the most significant distortions in a PPG system that lead to inaccurate measurement of the physiological parameters
~\citep{r35}.  Pressure disturbances arise from the contact between PPG sensor and skin/body area where PPG sensor is deployed~\citep{r35}. The arterial geometry of measurement site is changed due to pressure applied to the skin by the sensor cabinet. The pressure applied to the skin due to the placement of PPG sensor lead to undesirable changes in AC component of
the reflected PPG signal~\citep{r35}. Accurate estimation of heart rate from PPG signal corrupted by motion-artifacts has been  a challenging  task using time-domain
 as well as frequency-domain algorithms~\citep{r10}. The impact of various environmental distortions on a PPG signal quality is well studied in~\citep{r32} for infrared, red and green LEDs. The HR estimates obtained from green LED was found to be more accurate than those found with infrared and red light. The location of PPG sensor also impacts the detection quality due to the fact that the body's sweat rate and temperature vary at various locations. The contact between the skin and the sensor, movement in wearer's body part with the sensor, breathing, and physical activity degrade the quality of acquired PPG signal~\citep{r14}.
\subsection{Motion-Artifact Reduction in Wearable PPG Signal}
Many techniques have been proposed for reducing the motion-artifacts
in a PPG signal. Use of time-domain and frequency-domain independent component analysis (ICA) has been suggested, but it has two disadvantages. Firstly, the assumption of statistical independence between the PPG signal, and the motion-artifact does not hold in all cases. ICA required two acquisitions of motion corrupted PPG signals that provides additional burden on wearable PPG devices. Consequently, it requires multiple PPG sensors that might not be suitable for small wearable devices~\citep{r10},~\citep{r36}. Various adaptive signal processing techniques have been developed that use a reference signal to reduce the motion-artifacts. These techniques are not useful for everyday activities due to difficulty in estimating the  appropriate reference signal for such cases~\citep{r10}. \\
%
\begin{figure}[!t]
\centering
\includegraphics[width=0.5\textwidth]{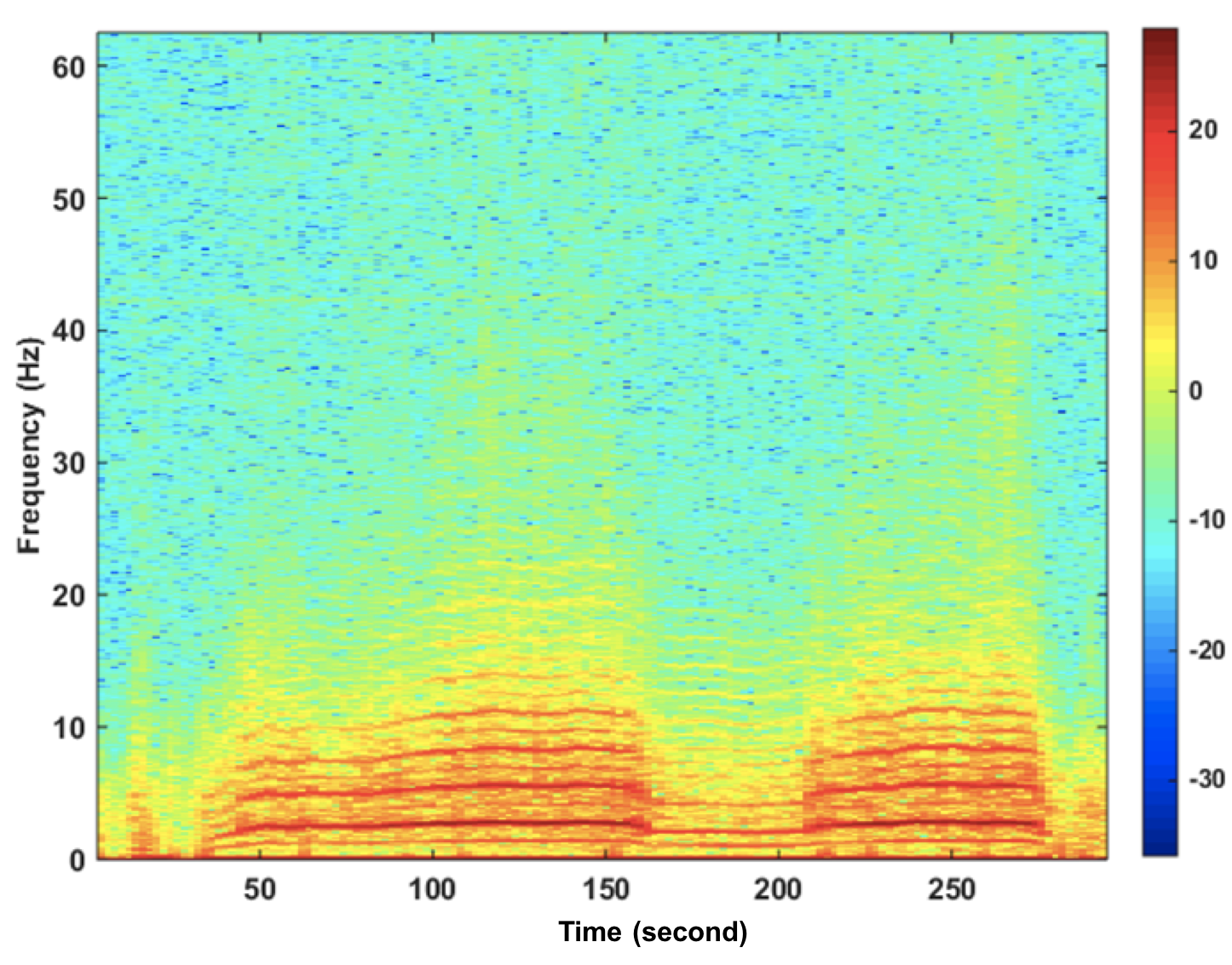}
\caption{The STFT of the accelerometer signal with a 2048-point fast Fourier transform (FFT) applied on 8-second time-windows where the successive windows have a  6-second overlap. The spectral amplitudes are quite significant upto about 12 Hz. Individual harmonics with fundamental frequency in the range of 1-3 Hz are evident. The complete acceleration signal of DATA05TYPE02 dataset was used for generating this figure~\citep{r10}.}
\label{fig_acc_spec}
 \end{figure}
Typically, the wristband systems such as smartwatches and fitness bands have an accelerometer that simultaneously records the acceleration signal~\citep{new1}. Other techniques being used for the motion-artifact and distortion reduction in PPG signal include the spectrum subtraction (subtracts the spectrum of the acceleration signal from the spectrum of the PPG signal)~\citep{r26}, adaptive filtering~\citep{r27}, higher-order statistics~\citep{r34}, wavelet transforms~\citep{r28}, empirical mode decomposition~\citep{r22}, time-frequency method described in~\citep{r20}, and the Kalman
 filtering~\citep{r37}. These methods are experimentally found to be effective only for small movements such as slow walking~\citep{r10}. Authors in~\citep{r10} suggested a generic and flexible framework, {TROIKA}, that is signal decomposi{T}ion for denoising, sparse
signal {R}ec{O}nstruct{I}on for high-resolution spectrum estimation, and spectral pea{K} tr{A}cking with verification. {TROIKA} was validated for HR estimation using wrist-type PPG signals while wearer runs at various speeds on a treadmill. Author proposed JOSS, that is, {JO}int {S}parse {S}pectrum reconstruction for accurate HR estimation using wrist-type PPG signal~\citep{zhang2015photoplethysmography}. It jointly estimated the spectra of the PPG signal and the acceleration signals. It is based on the multiple measurement vector (MMV) model for sparse signal recovery. A common sparsity constraint on the spectral coefficients helps in identification and removal of the spectral peaks corresponding to the motion-artifacts in the PPG spectrum. {JOSS} uses MMV model for sparse reconstruction, unlike TROIKA based on the single measurement vector (SMV) model~\citep{r10}. The {JOSS}~\citep{zhang2015photoplethysmography} exploits the common structures present in the spectrum of PPG signal and the spectrum of the acceleration signal and had shown better performance than the  {TROIKA} algorithm~\citep{r10}. A method for HR estimation based on Wiener Filtering and the Phase Vocoder ({WFPV}) was proposed in~\citep{temko2015estimation}.
Authors evaluated WFPV and concluded that it performed better than JOSS on average. The WFPV algorithm uses the accelerometer signal to estimate the motion-artifacts and later use a Wiener filer to attenuate the components of motion-artifacts in the corrupted PPG signal. The phase vocoder improved the resolution of estimation of dominant frequencies. Authors developed a method consisting of four stages namely, wavelet-based denoising, acceleration-based denoising, frequency-based heart rate estimation and finally a post-processing stage. This method  was found to be robust to motion-artifacts that occur during sports and rehabilitation~\citep{mullan2015unobtrusive}. An algorithm based on time-varying spectral filtering (named {SpaMA}) was proposed for accurate estimation of heart rate from PPG signals corrupted with motion-artifacts. Authors tested this approach over various datasets that were collected during various activities of daily life using wrist-band type PPG system~\citep{salehizadeh2015novel}.
\section{Materials and Methods}
In this section, we will describe the dataset, and discuss the motivation for development of harmonic sum (HSUM) models for HR estimation. Later, we will describe the mathematical derivations of the harmonic sum (HSUM) models based algorithm for HR estimation. We proposed a harmonic sum (HSUM) model for the measured acceleration signal and a joint HSUM model for the PPG signal corrupted with motion-artifacts. First, we perform an exploratory analysis of the signals that motivated the development of proposed algorithm. We evaluated the performance of HSUM algorithm on IEEE SP cup dataset. Later, we did a comparative analysis of HSUM with four methods that were recently developed namely {TROIKA}~\citep{r10}, {JOSS}~\citep{zhang2015photoplethysmography}, {WFPV}~\citep{temko2015estimation}, and {SpaMA}~\citep{salehizadeh2015novel}. 
\subsection{Datasets}
The scenarios used for acquisition of the IEEE SP cup data is described in ~\citep{r10}. The dataset consists of 12 motion affected PPG signals obtained from individuals while running on treadmill. It had dual-channel PPG signal along with simultaneously acquired ECG signal and three-axis acceleration signals. We found that for the proposed method using just one of the PPG channels was sufficient for heart rate extraction. We used the second channel for results discussed in this paper. The data was collected using a wrist-type PPG sensor while the wearer ran on a treadmill with increasing and decreasing
 speed for 5 minutes. The PPG signal, the accelerometer signal, and the  ECG signal were simultaneously recorded from 12 male subjects in the age range
 of 15-18 years. The wristband had a pulse oximeter with a green LED of wavelength 515 nm along with embedded accelerometers for acquisition of the
 PPG and the accelerometer signal. Wet ECG sensors were used to simultaneously collect the ECG data from the chest. The PPG, ECG and the
 accelerometer signals were sampled at 125 Hz. The acquired signals were sent to a nearby computer using  Bluetooth. The data were collected while
 the subjects walked or ran on a treadmill starting from rest to high speed before coming to rest again. Starting at a speed of 1-2 km/hour (kmph) for 30
 seconds, the speed was increased to 6-8 kmph for one minute followed by doubling the rate to 12-15 kmph for another one minute. For next two
 minutes, the same cycle is repeated, i.e., starting at speed of 6-8 kmph followed by 12-15 kmph. Finally, the subject walks at a speed of 1-2 kmph for
 30 seconds before coming to rest. 
 
 The ground-truth heart rate manually computed using the ECG signal were shipped with the dataset. The ground-truth HR for each overlapping time-window was computed by counting the number of cardiac cycles (H) and the duration (D) in seconds~\citep{r10}. The heart rate in beats per minute (BPM) is given by 
 \begin{equation}
 HR = \frac{60H}{D}
 \label{eqn_bpm}
 \end{equation}
 We did not use any algorithm for the estimation of heart rate (HR) from the ECG signal as it may cause estimation errors. We just used the provided ground-truth. The average absolute error $\xi_{HR}$ in HR estimates over N time-windows is defined as
 \begin{equation}
 \xi_{HR} = \frac{1}{N} \sum_{i=1}^N |HR[i]- \hat{HR[i]}|
 \label{eqn_bpm_error}
 \end{equation}
 where $HR[i]$ and $\hat{HR[i]}$ were the ground-truth and estimated HR value for the i-th time-window, respectively. 
\subsection{Motivations}
The signal acquired using a wrist-band worn by a person running on treadmill or similar intense physical exercise is severely corrupted with motion-artifacts.  Estimating the heart rate  from such a  PPG signal is challenging due to two facts. Firstly, the motion-artifacts are stronger than the heart-beat component in the PPG signal at several instances. Secondly, the spectrum of the heart-beat signal is close to the frequency range of the motion-artifact complicating the matter further. 

Figure~\ref{fig_time} shows an example of a PPG signal corrupted by the motion-artifacts and a simultaneously measured accelerometer signal. The quasi-periodicity in the accelerometer signal, shown in the bottom panel of the Figure~\ref{fig_time}, is quite evident. It contained the information about the motion-artifacts. Figure~\ref{fig_acc_spec} shows the Short-time Fourier Transform (STFT) of the accelerometer signal. The STFT (also known as a spectrogram) was obtained by 2048-point FFTs computed over 8-second time-windows with 6-second overlap between successive windows. The STFT shows a strong fundamental frequency component around 1 Hz along with several  higher harmonics of moderate intensity. 
\begin{figure}[!t]
\centering
\includegraphics[width=0.5\textwidth]{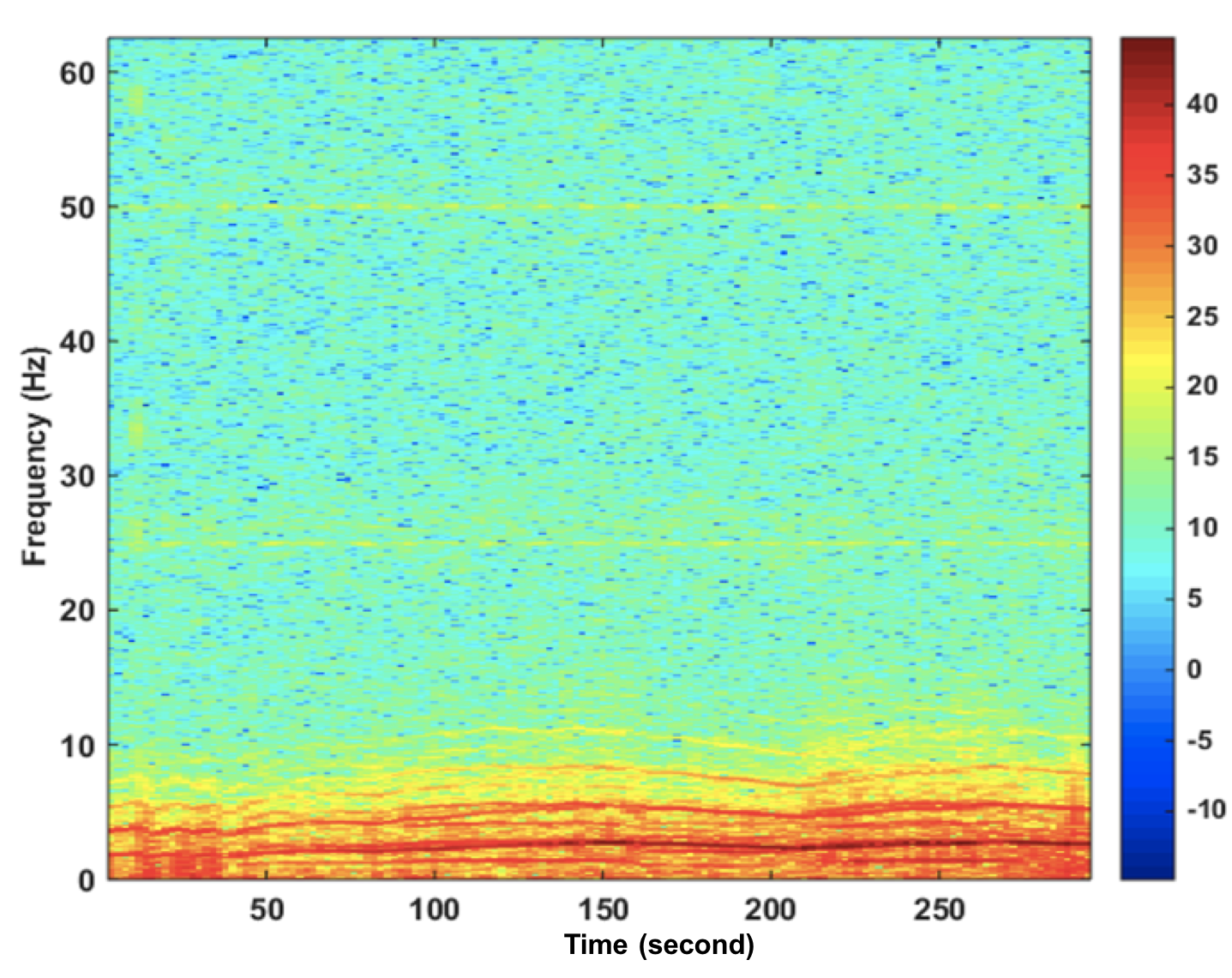}
\caption{The STFT of a PPG signal with 2048-point FFTs. The window sizes and overlap are the same as in Figure~\ref{fig_acc_spec}. The spectrum is dominant till about 6 Hz. Because of the presence of two sets of harmonics in the PPG signal the frequency tracks in this STFT are not as clean as in  Figure~\ref{fig_acc_spec}. It shows that the accelerometer signal and the heart rhythm signal have some overlapping spectral regions. The complete PPG signal in DATA05TYPE02 dataset of IEEE SP cup was used for generating this figure~\citep{r10}.}
\label{fig_ppg_spec}
\end{figure}
\begin{figure*}[!t]
\centering
\includegraphics[width=0.99\textwidth]{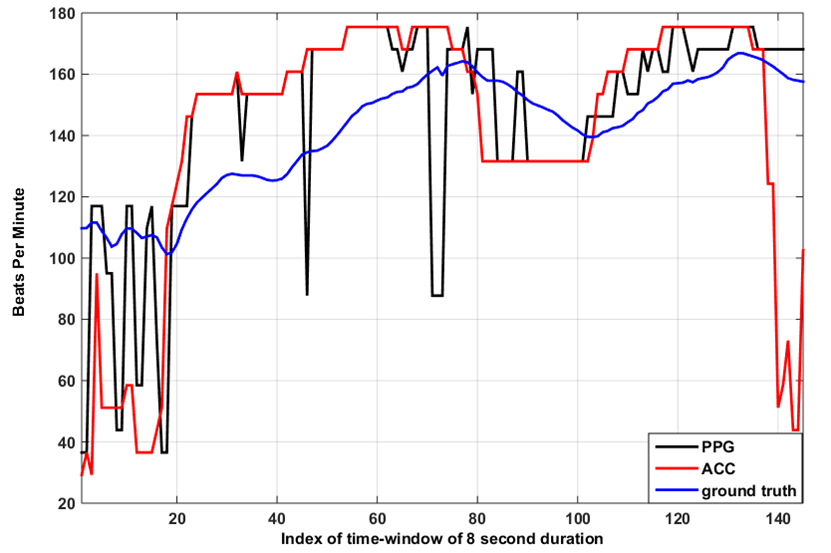}
\caption{The purpose of this figure is to point out that picking the largest peak  of the STFT magnitude of the measured PPG signal in each time window
 gives  incorrect estimates of the instantaneous  heart rate. The STFT of a  signal was obtained with 2048-point FFTs computed over   8-second time-windows where the successive windows had 6-second overlap. The frequency location  of the peak of
 the magnitudes of the STFT in each time  window were obtained
 and multiplied by 60 to give an estimate of the heart rate in  beats per minute.
 The heart rate in beats per minute for the  PPG signal is denoted by the black solid
 line. The ground-truth heart rate (HR) obtained from the simultaneously acquired ECG
 signal is shown by the blue line. As can be seen, the heart rate
 estimates obtained from the PPG STFT  peaks significantly deviate from 
 the ground-truth. Also shown for comparison purposes, are the peak  locations
 obtained from the accelerometer signal's STFT (red line). Notice that in many time windows, the estimates given by the PPG signal and the peak-magnitude locations of the  accelerometer signal's FFT coincide. This shows that often  the spectrum of motion-artifacts overlap with that of the heart-beat component of the PPG signal.}
\label{fig_stft}
\end{figure*}
\begin{figure*}[!t]
\centering		
\includegraphics[width=0.99\textwidth]{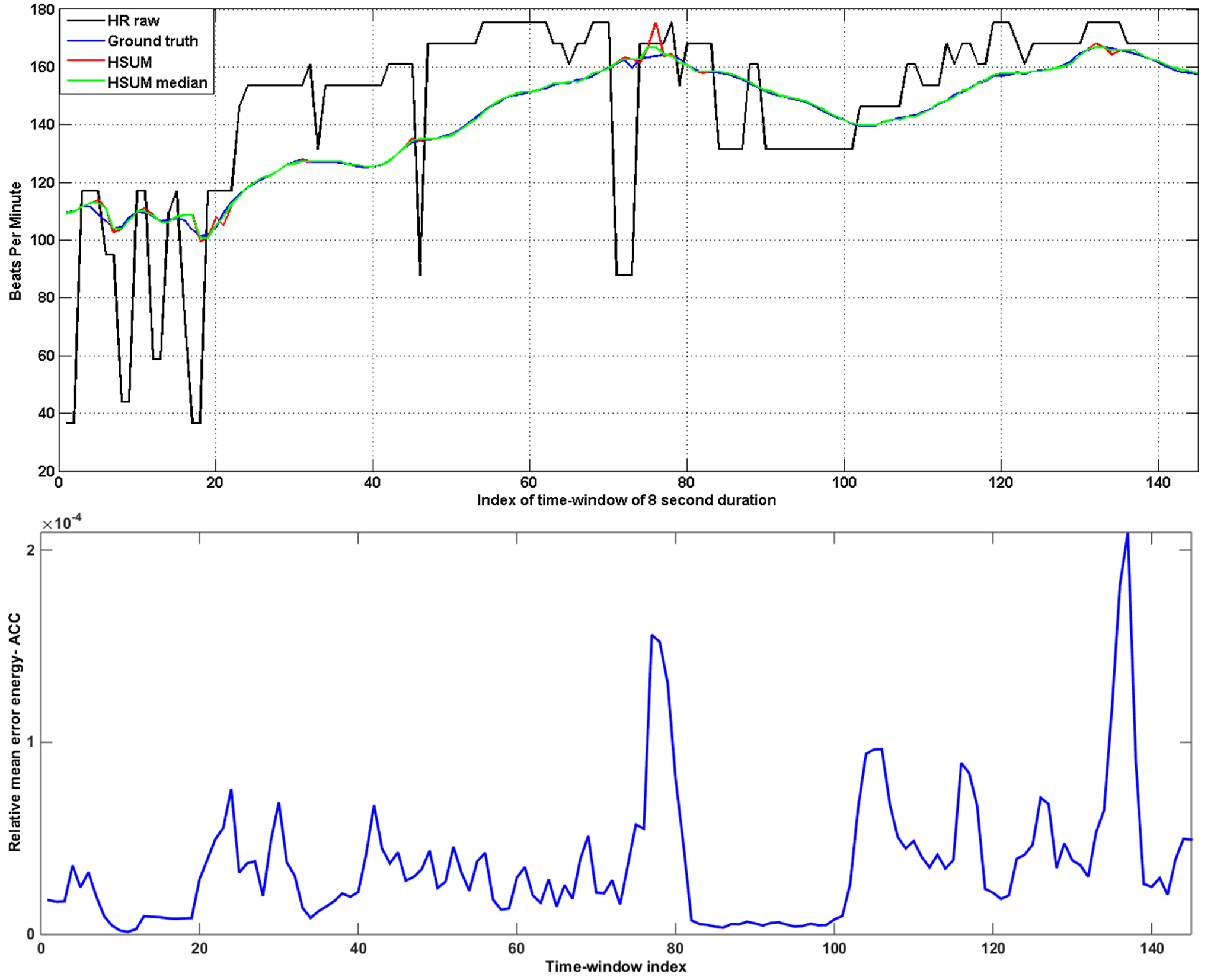}	
\caption{Top Panel: Comparison of heart rate estimates obtained using the HSUM model and the ground-truth heart rate. Time-windows of 8-second duration with 6-second overlap between the successive windows were used. The 'HR raw' are the HR obtained directly from the peak locations of the spectrogram of the measured PPG signal (as in Figure~\ref{fig_stft}). Harmonic sum-based estimates are almost the same as the ground-truth heart rate estimates except at a couple of  points. The mean absolute error is 0.6970 beats per minute (BPM). The data used was DATA05TYPE02 from~\citep{r10}. The 'HSUM median' line corresponds to a 3-point median filtered estimates of harmonic sum-based method. It slightly improves the estimates obtained from harmonic sum (HSUM) modeling. Bottom Panel: Shows the relative mean error energy (obtained by dividing the mean squared error(SE) by the energy of the acceleration signal) over successive time-windows. Notice that in general, larger relative mean error energy corresponds to greater deviation of the estimated heart rate from ground-truth.}
\label{fig_hr_all}
\end{figure*}
Figure~\ref{fig_time} (top panel) shows the PPG signal. Clearly the PPG signal shows significant envelope fluctuations. A cursory examination of the waveform shows that the envelope fluctuations have a frequency of roughly 0.2 to 0.4 Hz. This is due to the fact that the harmonic components of the heart rhythm interacting with the harmonic components of the motion related  signal, i.e., dominant components  of the two periodic signals are quite close  to each other in frequency thereby producing a 'beat signal' envelope. The Figure~\ref{fig_ppg_spec} shows the STFT of the PPG signal computed using the same parameters as in Figure~\ref{fig_acc_spec}. The PPG signal has less number of significant  harmonics when compared to the accelerometer signal. Further, in the STFT, we notice that  in the   low frequency region there is significant interaction between the two quasi-periodic signals. Hence unlike the STFT for the accelerometer signal, the  frequency tracks are somewhat jumbled. Thus, it is not possible to resolve the individual harmonic components of the two periodic signals using standard Fourier transform unless the time window is made wider or sampling rate is increased. However, the time window can not be made much wider because then  the heart rhythm signal and the motion-artifact related signal might change their rates within that wider window thereby smearing the frequency tracks. This is a classical problem in time-frequency analysis methods like the STFT. To increase resolution of the STFT, the sampling rate could
also be increased.

Figure~\ref{fig_stft} shows the problems with estimating the heart rate from the locations of the peaks in the Short-Time Fourier Transform(STFT) of the PPG signal.
As an exploratory step, we computed the heart rates   from the STFT magnitude of the motion-artifact corrupted PPG signal. The PPG signal was divided into overlapping windows of 8-second duration with 6-second overlap between successive windows. Each of the (Hanning) windowed segment of PPG signal was processed with a 2048-point fast Fourier transform (FFT).  The frequency location of the largest peak in the  magnitude of the STFT  was  used to obtain the heart rate estimate for each window.  This frequency location  in Hz is multiplied by 60 to get the heart rate in beats per minute (BPM). This heart rate obtained for the  PPG signal  is plotted for each time window  in Figure~\ref{fig_stft} (solid black  line).
 The ground-truth heart rate  obtained from the simultaneously acquired ECG signal is shown by the blue line. As can be seen, the heart rate estimates obtained from the PPG signal's  STFT-magnitude  peaks' locations  wildly fluctuate and also significantly deviate from  the ground-truth.  Also shown for comparison purposes, are the frequency locations of the magnitude peak of the STFT (converted to BPM) of the accelerometer signal (red line). Notice that in many time windows, the estimates given by the PPG signal (black line) and the BPM values corresponding to the  peak-magnitude locations of the  accelerometer signal (red line) coincide.  This shows that often  the  motion-artifacts are much stronger than the heart-beat component in the
 PPG signal. In other words, the motion-artifacts often  dominate the PPG signal, and so they need to be some how suppressed to the extent possible from the measured PPG signal before heart rate estimation.
To overcome some of the above mentioned problems, we propose a novel approach. Since both the heart rhythm signal and the motion-artifact related signal appear to be quasi periodic in nature,  we propose to use  a truncated Fourier series to model these signals over short time windows. Such models have been previously used, for example, in processing voiced speech sounds (vowels)~\citep{r9}. We propose the following strategy. First model the accelerometer signal using a truncated Fourier series (HSUM), and estimate its fundamental frequency. Then,  since the PPG signal is a composite of the heart rate signal and the motion-artifact related signal, we fit a sum of two different truncated Fourier series models (joint HSUM) to the PPG signal. One of the harmonic sums corresponds to the heart-beat component in PPG and the other models the motion-artifact. However, the fundamental frequency of the motion-artifact has already been determined from the accelerometer signal in the first step. Using this estimate, in the next step we  estimate the fundamental frequency of the other periodic component that obtains the heart rate. 
\begin{figure*}[!t]
\centering		
\includegraphics[width=0.99\textwidth]{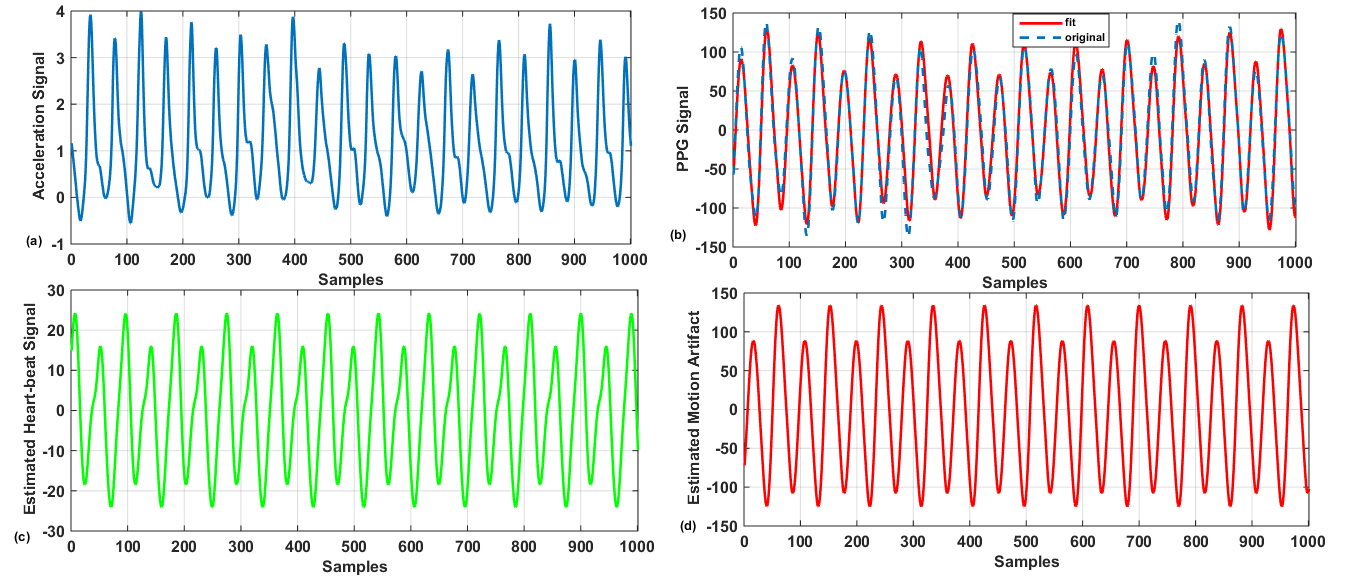}
\caption{(a) An example of the accelerometer signal for a 8-second time-window, (b) A sample PPG signal and its joint HSUM model fit for the same window, (c) Heart-beat component of the PPG signal shown in figure (b) obtained by using Equation~\ref{eqn_xheartbeat}, (d) Motion-artifact component of the PPG signal using Equation~\ref{eqn_artifact}. This segment is taken from DATA05TYPE02 dataset~\citep{r10} in a time interval when the individual is running at the rate of 12 kmph. It can be seen from high magnitude of acceleration in this segment. Since the HSUM model is fitted over a 8-second window over which the acceleration signal and PPG signal are quasi-periodic, similar figures would be obtained for 6 kmph and 15 kmph for example. Higher speed of motion cause higher corruption of PPG signals with motion artifacts.}
\label{fig_heart}
\end{figure*}
\subsection{Harmonic Sum (HSUM) for the Accelerometer Signal}
Let us first consider the simpler problem of modeling the accelerometer signal, since it is assumed to consist of only one quasi-periodic signal. Let us assume that we have $N_{a}$ samples
 of the accelerometer signal. It is modeled as a sum of a DC component $a_{0}$ and $M_{a}$   sines and cosines with frequencies that  are integer multiples of the fundamental frequency $f_a$ Hz.  The amplitudes are denoted by $a_k$ and $b_k$.  This HSUM model is denoted by $\mathbf{\hat{x}_{a}}$.
 \begin{equation}
\hat{x}_{a}[n]= a_{0} + \sum_{k=1}^{M_{a}}a_{k}\cos\left({\frac{2 \pi knf_{a}}{f_s}}\right) + \sum_{k=1}^{M_{a}} b_{k}\sin\left({\frac{2 \pi knf_{a}}{f_s}}\right)
\label{eqn_hsum_acc}
\end{equation}
The unknown amplitudes and fundamental frequency $f_a$ in  the above equation are estimated by minimizing the squared error (SE) between original signal $\mathbf{x_{a}}$ and the model $\mathbf{\hat{x}_{a}}$, 
\begin{equation}
SE =  \sum_{n=0}^{N_{a}-1} \left( \hat{x}_{a}[n] - x_{a}[n] \right)^2.
\label{eqn_se_acc}
\end{equation}
We can vectorize Equations~\ref{eqn_hsum_acc} and~\ref{eqn_se_acc} by writing
\begin{equation}
\mathbf{x_{a}}= \left( x_{a}[0], x_{a}[1],...,x_{a}[N_{a}-1] \right)^T,
\label{eqn_xa}
\end{equation}
\begin{equation}
\mathbf{\hat{x}_{a}} =  \left( \hat{x}_{a}[0], \hat{x}_{a}[1],...,\hat{x}_{a}[N_{a}-1] \right)^T,
\label{eqn_hxa}
\end{equation}
\begin{equation}
\mathbf{a_{a}}= \left( a_{0},a_{1},..,a_{M_{a}}, b_{1},..,b_{M_{a}} \right)^T,
\label{eqn_aa}
\end{equation}
and a matrix $\mathbf{W_{a}}$ defined as follows.
\begin{equation}
W_{a}[k,l]= 
\begin{cases}
1 &\text{k=0; l=1,..,2$M_{a}$}\\ 
\text{cos}  \left( \frac{2\pi klf_{a}}{f_s} \right), &\text{k=1,..,$N_{a}$-$1; l=1$,..,$M_{a}$}\\
\text{sin}\left( \frac{2\pi k(l-M_{a})f_{a}}{f_s} \right),     &\text{k=1,.,$N_{a}$-1;l=$M_{a}+1$,.,$2M_{a}$}\\
\end{cases}
\label{eqn_wa}
\end{equation}
where $W_{a}[k,l]$ stands for the $(k,l)^{th}$ element of the $\mathbf{W_{a}}$ matrix. The Equation~\ref{eqn_hsum_acc} can be rewritten in vectorized form as
\begin{equation}
\mathbf{x_{a}}= \mathbf{W_{a}} \mathbf{a_{a}}.
\label{eqn_vacc}
\end{equation}
The Equation~\ref{eqn_se_acc} can then be rewritten in vector form as
\begin{equation}
SE= || \mathbf{x_{a}} - \mathbf{W_{a}} \mathbf{a_{a}} ||_2^{2}.
\label{eqn_vse}
\end{equation}
Minimizing SE by choosing the unknown parameters is a bi-linear least squares problem, since both $\mathbf{W_{a}}$ matrix and $\mathbf{a_{a}}$ are unknown. But  it  can be simplified as follows. The squared error in Equation~\ref{eqn_vse} can be minimized by a standard least squares method (See for example,~\citep{r9}) if the frequency $f_a$ (and hence $\mathbf{W_{a}}$)  is known. For a given frequency $f_{a}$, we can form the matrix  $\mathbf{W_{a}}$  as per  Equation~\ref{eqn_wa}. Then the amplitude vector $\mathbf{a_{a}}$ that minimizes the squared error is  given by~\citep{r9}
\begin{equation}
\mathbf{a_{a}} = \left( \mathbf{W_{a}}^T\mathbf{W_{a}} \right)^{-1} \mathbf{W_{a}}^T \mathbf{x_{a}}.
\label{eqn_vaa}
\end{equation}
Substituting the above expression for  $\mathbf{a_{a}}$ back in Equation~\ref{eqn_vse}, we can rewrite the squared error (SE) as 
\begin{equation}
SE= || \mathbf{x_{a}} - \mathbf{W_{a}} \left( \mathbf{W_{a}}^T\mathbf{W_{a}} \right)^{-1} \mathbf{W_{a}}^T \mathbf{x_{a}} ||_2^{2}.
\label{eqn_sea2}
\end{equation}
We shall  define  a projection matrix $\mathbf{P_{a}}$,  as follows 
\begin{equation}
\mathbf{P_{a}}= \mathbf{W_{a}} \left(\mathbf{W_{a}}^T \mathbf{W_{a}}\right)^{-1} \mathbf{W_{a}}^T.
\label{eqn_pa}
\end{equation}
By noting that the matrices, $\mathbf{P_{a}}$ and $\mathbf{I - P_{a}}$	are idempotent, we can rewrite the squared error in Equation~\ref{eqn_sea2} as follows~\citep{r9}.
\begin{equation}
SE =  \mathbf{x_{a}}^{T} \left( \mathbf{I} -\mathbf{P_{a}}\right) \mathbf{x_{a}}.
\label{eqn_sea3}
\end{equation}
Note that the squared error, SE, explicitly depends only on the unknown frequency $f_{a}$. We can either minimize the expression in Equation~\ref{eqn_sea3}, or equivalently, maximize $\mathbf{x_{a}}^{T} \mathbf{P_{a}}\mathbf{x_{a}}$  by picking the best $f_a$. Minimization of  SE can be achieved  by searching over a grid covering the  range of expected frequency values of $f_{a}$. Since we know that the accelerometer signal has a fundamental frequency in the range of 1 to 3 Hz, we can use a  grid search over this range of frequencies with a step size of, say, 0.01 Hz. By minimizing the SE (Equation~\ref{eqn_sea3}) we estimate the fundamental frequency of the  HSUM model for each overlapping window. In our processing algorithms we used
 overlapping windows of 8-second duration with 6-second overlap between  successive windows. Equation~\ref{eqn_vaa} can then be used to compute the amplitudes of all harmonics for each window. The optimum frequency estimate  computed here  is used for estimating the parameters of a joint   HSUM model for the PPG signal as described in the next section.
\subsection{Joint Harmonic Sum (HSUM) model for the PPG Signal}
The PPG signal acquired during daily activities is composed of the 
heart-beat signal  and  dominant motion-artifacts induced by the physical movements of the user. Unfortunately, we do not know how exactly the physical movements of the subject affect the PPG signal. Although in the previous subsection we obtained a model fit to the accelerometer data and estimated the fundamental frequency $f_a$ and the amplitudes  $\mathbf{a_{a}}$ we are uncertain as to  how the individual harmonic's amplitudes affect the PPG data. But we may  hypothesize that the artifacts induced by  physical movements  in the PPG signal have the same fundamental frequency as that of the  accelerometer signal. If this were true, then we can  use the fundamental frequency estimate $f_a$ obtained in the previous subsection (but not the amplitudes $\mathbf{a_{a}}$)  to help mitigate the effects of the motion-artifacts on  the PPG signal. Our experimental results below seem to validate this hypothesis.
The harmonic sum (HSUM) model for the PPG signal consists of a  sum of two truncated harmonic series with different fundamental frequencies, $f_{a}$ for the motion-artifact component, and $f_{h}$ for the heart-beat component. The value for $f_{a}$ that gives minimum squared error (SE) for accelerometer signal fit is taken as the optimum fundamental  frequency of the motion-artifact (and renamed as $f_{oa}$ for ease of use). The signal model for the PPG signal is  then given by the Equation,
\begin{multline}
\hat{x}_{p}[n]= a'_{0} + \sum_{k=1}^{M_{a}} a'_{k} \cos\left({\frac{2 \pi knf_{oa}}{f_s}}\right) +\sum_{k=1}^{M_{a}} b'_{k} \sin\left({\frac{2 \pi knf_{oa}}{f_s}}\right)\\
+ \sum_{\kappa=1}^{M_{h}} c_{\kappa} \cos\left({\frac{2 \pi \kappa nf_{h}}{f_s}}\right) +  \sum_{\kappa=1}^{M_{h}}d_{\kappa} \sin\left({\frac{2 \pi \kappa nf_{h}}{f_s}}\right).
\label{eqn_xp}
\end{multline}
The model represented by the Equation~\ref{eqn_xp} can be vectorized as in the previous subsection as follows.
\begin{equation}
\mathbf{\hat{x}_{p}}= \mathbf{W_{oa}} \mathbf{\bar{a}_{a}} + \mathbf{W_{h}} \mathbf{c_{h}} 
\label{eqn_vxp}
\end{equation}
where the weight matrix, $\mathbf{W_{h}}$, and  the
amplitude vector of the  heart-beat component of the  PPG signal, $\mathbf{c_{h}}$ are defined as follows.
\begin{equation}
\mathbf{c_{h}} =  \left(c_{1},..,c_{M_{h}}, d_{1},..,d_{M_{h}} \right)^T
\label{eqn_c}
\end{equation}
and 
\begin{equation}
W_{h}[k,l]= 
\begin{cases}
\text{cos}  \left( \frac{2\pi klf_{h}}{f_s} \right), &\text{k=1,..,$N_{h}$-$1; l=1$,..,$M_{h}$}\\
\text{sin}\left( \frac{2\pi k(l-M_{h})f_{h}}{f_s} \right),     &\text{k=1,.,$N_{h}$-1;l=$M_{h}+1$,.,$2M_{h}$}\\
\end{cases}
\label{eqn_wh}
\end{equation}
respectively. Here, $W_{h}[k,l]$ stands for the $(k,l)^{th}$ element of the $\mathbf{W_{h}}$ matrix. The known fundamental frequency $f_{oa}$ is  used to synthesize the optimum weight matrix, $\mathbf{W_{oa}}$, for the acceleration signal using Equation~\ref{eqn_wa} (i.e., substituting $f_{oa}$ in place of $f_a$). The amplitudes of motion-artifact component, $\mathbf{\bar{a}_{a}}$ are specified in vector form as follows.
\begin{equation}
\mathbf{\bar{a}_{a}} =  \left(a'_{0},a'_{1},..,a'_{M_{a}}, b'_{1},..,b'_{M_{a}} \right)^T,
\label{eqn_abar}
\end{equation}
The combined amplitude vector of both motion-artifact component and heart-beat component of the PPG signal is given by
\begin{equation}
\mathbf{a_{c}} =  [\mathbf{\bar{a}_{a}}, \mathbf{c_{h}}]^{T}.
\label{eqn_vac}
\end{equation}\\
Similarly, by concatenating the weight matrices corresponding to heart-beat component, $\mathbf{W_{h}}$ and the motion-artifact related signal $\mathbf{W_{oa}}$, we get the combined weight matrix, $\mathbf{W_{c}}$, as 
\begin{equation}
\mathbf{W_{c}} = \left( \mathbf{W_{oa}} | \mathbf{W_{h}}\right).
\label{eqn_vwc}
\end{equation}
Now, analogous to Equation~\ref{eqn_vse}, we can write the squared error for the PPG signal as follows.
\begin{equation}
SE_p= || \mathbf{x_{p}} - \mathbf{W_{c}} \mathbf{a_{c}} ||_2^{2}.
\label{eqn_sep}
\end{equation}
where $\mathbf{x_{p}}$ stands for the observed PPG signal vector. Following the same steps as in the case of accelerometer modeling, the combined amplitude vector that minimizes $SE_p$ for a given $f_h$ is given by
\begin{equation}
\mathbf{a_{c}} = \left( \mathbf{W_{c}^T} \mathbf{W_{c}} \right)^{-1} \mathbf{W_{c}}^T \mathbf{x_{p}}.
\label{eqn_vac2}
\end{equation}
The projection matrix for joint HSUM model for PPG signal is given by
\begin{equation}
\mathbf{P_{p}}= \mathbf{W_{c}} \left(\mathbf{W_{c}}^T \mathbf{W_{c}}\right)^{-1} \mathbf{W_{c}}^T.
\label{eqn_vpp}
\end{equation}
The corresponding squared error, $SE_{p}$, in vector form is written as
\begin{equation}
SE_{p} =  \mathbf{x_{p}}^{T} \left( \mathbf{I} -\mathbf{P_{p}}\right) \mathbf{x_{p}}.
\label{eqn_sep2}
\end{equation}	
The frequency, $f_{oh}$, that gives the minimum $SE_p$  is the heart rate in Hz. We multiply it by 60 to get the heart rate  in beats per minute (BPM) as indicated in Figure~\ref{fig_hsum}. For finding the best $f_h$ (called $f_{oh}$), we used a grid search in the range of 0.5 to 3 Hz (in steps of 0.01 Hz). For
 each frequency in the grid, we compute corresponding weight matrix, $\mathbf{W_{h}}$ with a different order $M_{h}$ . $\mathbf{W_{oa}}$ is of course
 fixed. Once $f_{oh}$ is determined, then all the optimal amplitudes can be estimated using the expression in Equation~\ref{eqn_vac2}.  
 Using the joint HSUM model, we can now suppress the motion-artifacts present in the PPG signal leaving  behind the heart-beat component of the PPG signal.
We can reconstruct  the heart-beat related signal component $\hat{x}_{hb}[n]$  using the expression,
\begin{equation}
\hat{x}_{hb}[n]= 
\sum_{\kappa=1}^{M_{h}} c_{o\kappa} \cos\left({\frac{2 \pi \kappa nf_{oh}}{f_s}}\right)
+\sum_{\kappa=1}^{M_{h}}d_{o\kappa} \sin\left({\frac{2 \pi \kappa nf_{oh}}{f_s}}\right),
\label{eqn_xheartbeat}
\end{equation}
where $c_{o\kappa}$ and $d_{o\kappa}$ are the optimum amplitude estimates obtained from Equation~\ref{eqn_vac2}. 
Similarly, we can reconstruct the motion-artifact component of the PPG signal using the following expression,
\begin{equation}
\hat{x}_{a}[n]= 
\sum_{k=1}^{M_{a}} a'_{ok} \cos\left({\frac{2 \pi k nf_{oa}}{f_s}}\right) + \sum_{k=1}^{M_{a}}b'_{ok} \sin\left({\frac{2 \pi k nf_{oa}}{f_s}}\right),
\label{eqn_artifact}
\end{equation}
where $c_{ok}$ and $d_{ok}$ are the optimum amplitude estimates obtained from Equation~\ref{eqn_vac2}.
Figure~\ref{fig_heart} shows the application of HSUM to a 8-second time-window of the PPG signal and the accelerometer signal. The figure ~\ref{fig_heart} (a) shows the time-domain accelerometer signal for 8-second duration (1000 samples at sampling rate of 125 Hz). The quasi-periodic structure in the time-domain acceleration signal is evident from this figure.
 The Figure~\ref{fig_heart} (b) shows the PPG signal and its joint HSUM model fit for the same time-window. We can see that the joint HSUM model closely follows the PPG signal. Finally, Figure~\ref{fig_heart}(c) and~\ref{fig_heart}(d) shows the heart beat component of the PPG
 signal computed using Equation~\ref{eqn_xheartbeat}, and the motion-artifact component of 
the PPG signal obtained using Equation~\ref{eqn_artifact}. We can see that the amplitude of the heart-beat component of the PPG signal is significantly lower than amplitude of the motion-artifact component. The beauty of the harmonic sum (HSUM) model lies in the fact that it fit the frequency of both components, the heart-beat and motion-artifact, using the joint HSUM model.
We can summarize the proposed algorithm as follows. We first process the accelerometer data over a short time window using  the HSUM model in Equation~\ref{eqn_hsum_acc}. We minimize the squared error in Equation~\ref{eqn_se_acc} by finding the optimum fundamental frequency $f_{oa}$ valid over that time window. Then we process the PPG data over the same time window
using the joint HSUM model in Equation~\ref{eqn_xp}. We minimize the squared error in Equation~\ref{eqn_sep2} by finding the optimum fundamental frequency $f_{oh}$ while making use of the $f_{oa}$ value obtained in the first step. The $f_{oh}$ is finally converted to a heart rate estimate valid over that time window. We then repeat this process for each overlapping time-window. We now describe the results of applying this algorithm on IEEE SP cup data.
\section{Results and Discussions}
\begin{figure*}[!t]
\centering
\includegraphics[width=0.99\textwidth]{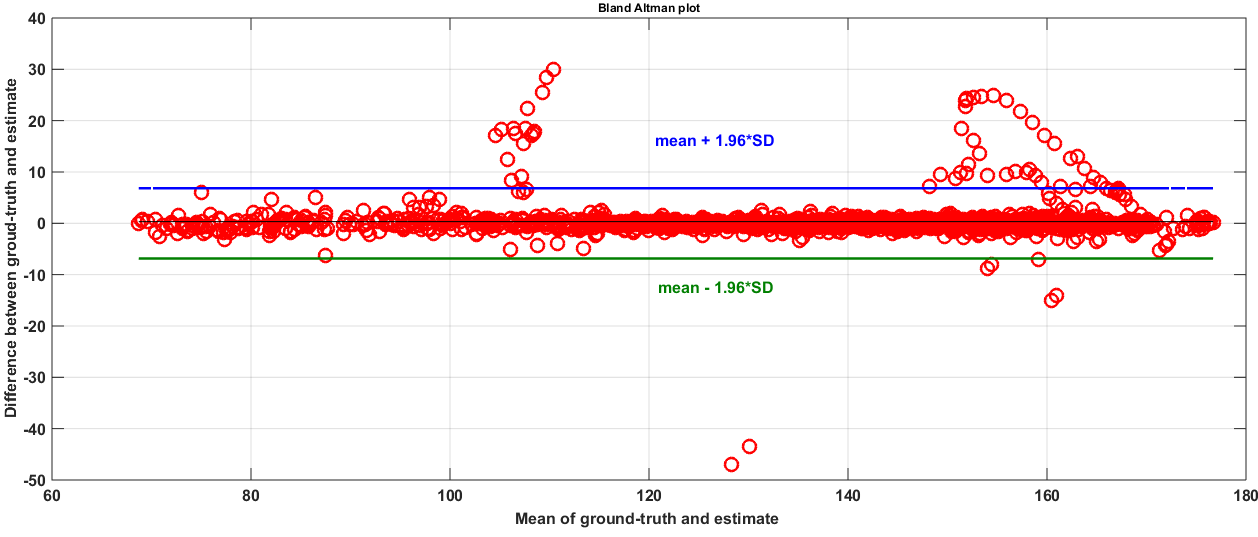}
\caption{The Bland-Altman plot between HR estimates obtained using the HSUM model with 3-point median filtering and the ground-truth from the ECG signal for all 12 data-sets in IEEE SP cup data~\citep{r10}. Strong agreement in HR estimates with the ground-truths is evident. Most of the points are inside the limit of agreement that is at $\pm$ 1.96 times the standard deviation~\citep{r25}.}
\label{fig_ba}
\end{figure*}
This section describes the experiments conducted to validate the efficacy of the proposed HSUM model for estimation of HR from the PPG signal corrupted with motion-artifacts. 
\begin{figure*}[!t]
\centering
\includegraphics[width=0.99\textwidth]{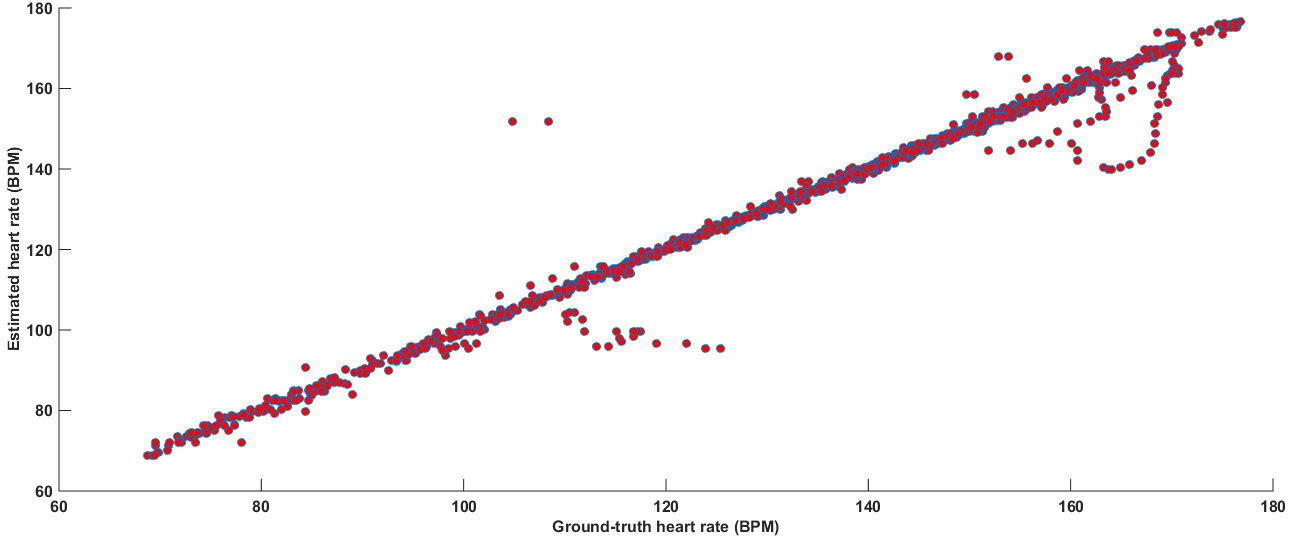}
\caption{The Pearson's correlation between HR estimates obtained using the HSUM model with 3-point median filtering and ground-truth from the  ECG for all 12 data-sets in IEEE SP cup data~\citep{r10}. The Pearson's correlation coefficient was 0.9974 and Spearman's rank coefficient was 0.9978. High values of these coefficients suggest that the estimated heart rates were almost same as the ground-truth.}
\label{fig_pear}
\end{figure*}
\subsection{Results}
\begin{table*}[!t]
\caption{Average absolute error(in beats per minute) for HR estimates obtained from the HSUM model, the HSUM model with 3-point median filtering compared with the HR estimates obtained from the STFT of the PPG signal, and using 3-point median filtering on HR estimates obtained from the STFT of the PPG signal. Also, the average absolute error (in BPM) using other four recently developed algorithms namely SpaMA~\citep{salehizadeh2015novel}, WFPV~\citep{temko2015estimation}, JOSS~\citep{zhang2015photoplethysmography}, and TROIKA~\citep{r10} is also given for comparison. These accuracies are taken from respective references. We can see that HSUM gives an improvement over these algorithms. The improved accuracy of HR estimates obtained using HSUM shows the applicability of the model to the given problem.}
\label{avg_error}
\centering
\begin{tabular}{*{13}{c}}
\hline
&S1&S2&S3&S4&S5&S6&S7&S8&S9&S10&S11&S12\\
\hline
HSUM \\
median&0.614&0.762&0.649 & 0.592 & 0.534 & 0.522&0.592&0.512& 0.412 &0.583&1.484&1.576 \\
\hline
SpaMA \\
~\citep{salehizadeh2015novel} & 1.23 & 1.59 & 0.57 & 0.44 & 0.47 & 0.61 & 0.54 & 0.40 & 0.40 & 2.63 & 0.64 & 1.20 \\
\hline
WFPV \\
~\citep{temko2015estimation} & 1.23 & 1.26 & 0.72 & 0.98 & 0.75 & 0.91 & 0.67 & 0.91 & 0.54 & 2.61 & 0.94 & 0.98 \\
\hline
JOSS \\
~\citep{zhang2015photoplethysmography} & 1.33 & 1.75 & 1.47 & 1.48 & 0.69 & 1.32 & 0.71 & 0.56 & 0.49 & 3.81 & 0.78 & 1.04 \\
\hline
TROIKA \\
~\citep{r10} & 2.87 & 2.75 & 1.91 & 2.25 & 1.69 & 3.16 & 1.72 & 1.83 & 1.58 & 4.00 & 1.96 & 3.33  \\
\hline
HSUM&0.756&0.917&0.948&1.185&0.697&0.609&0.873&0.594&0.525&0.754&1.495& 2.469\\		 
\hline
Short-time \\
Spectrum \\
Median&49.158&42.563&41.419& 28.521&10.559& 15.972&8.327&18.748&7.534&53.276&14.608&22.381\\
\hline	
Short-time \\
Spectrum&49.154&43.730&42.664&30.762&12.106&17.100&10.100& 19.487&7.927&55.569&15.690&25.892\\
\hline		
\end{tabular}
\end{table*}
\begin{table*}[!t]
\caption{Standard deviation in Error (in beats per minute) for HR estimates obtained from HSUM model, HSUM model with 3-point median filtering compared with the HR estimates obtained from the STFT of the PPG signal, and using 3-point median filtering on HR estimates obtained from the STFT of the PPG signal.}
\label{std_error}
\centering
\begin{tabular}{*{13}{c}}
\hline
& S1 & S2 & S3 & S4 & S5 & S6 & S7 & S8 & S9 & S10 & S11 & S12 \\
\hline
HSUM \\
median & 0.500 &0.966& 0.710& 0.636&0.777& 0.979&0.796& 0.470 &0.372&0.401&1.865&1.522\\
\hline
HSUM &0.604&1.201&1.326&1.735&1.274&1.133&1.307&0.623&0.671&0.522& 1.840&1.767\\		 
\hline
Short-time \\Spectrum \\ median & 45.121&29.640&32.538&39.154&15.528& 24.145&7.550&22.518&2.172&46.245&16.417&15.378\\	
\hline	
Short-time \\
Spectrum &44.469&30.404&32.654&41.3033&20.292&25.484&14.167& 22.601&4.404&48.669&19.183&23.484\\	
\hline		
\end{tabular}
\end{table*}
Figure~\ref{fig_hr_all} compares the heart rate estimates obtained using the  HSUM model, the HSUM followed by 3-point median filtering, frequency locations of the peak of the STFT magnitude of  the  PPG signal with the
 ground-truth. The time-windows of 8-second duration with 6-second overlap between successive windows were used. It shows that the heart rate estimates obtained
 using the  HSUM model are almost the same as the ground-truth heart rate(HR) estimates except at a few points. The mean absolute error was 0.6970 beats per minute (BPM) for the dataset, DATA05TYPE02 (from~\citep{r10}). We used a 17-th order HSUM model ($M_a=17$) for accelerometer signal and 7-th order HSUM model ($M_h=7$) for the  heart-beat component of the PPG signal. We used a 3-point median filter to refine the HR estimates obtained from the
 HSUM model. The 3-point median filter incorporates HR estimates from previous and next windows. Clearly, maximum refinement of HR estimates is seen at the point of highest deviation from the ground-truth~\ref{fig_hr_all}. Average absolute error in HR estimates with HSUM model was 0.9852 BPM with standard deviation of 1.1670 BPM for the complete dataset. On using 3-point median filter, we got an average absolute error of 0.7359 BPM with standard deviation of 0.8328 BPM. On an online scheme where we the future frame (next 2 second) of PPG and acceleration signals are not available, the median filtering can be skipped. It is to be noted that HSUM is effective in accurate modeling of PPG and acceleration signals over short overlapping windows. HSUM gives accurate heart rates without median filtering. However, authors include median filtering for cases where we can have access to next 2 seconds of signals (or equivalently we are in a offline scheme where short delays are acceptable). With 2 second delay, we can refine the HR estimates by incorporating the context (that is done by median filtering). On the other hand, average absolute error in HR estimates obtained from STFT is 27.5152 BPM with standard deviation of 27.2596 BPM. Using median filter on HR estimates from short-time spectrum we get average absolute error of 26.0886 BPM with standard deviation of 24.7005 BPM. The error in HR estimates obtained from the STFT is very large because motion-artifacts have corrupted the PPG signal significantly due to considerable motion of the subject's hand while running on the treadmill. The Figure~\ref{fig_stft} shows the HR estimates obtained from the STFT magnitude of the PPG signal as well as the accelerometer
 signal. Large error in HR estimates is evident from this figure. 
 
Table~\ref{avg_error} depicts the average absolute error (in beats per minute) in HR estimates obtained from the HSUM model, the HSUM model with 3-point median filtering compared with the HR estimates obtained from the STFT of the PPG signal, and using 3-point median filtering on HR estimates obtained from the STFT of the PPG signal. Also, the average absolute error (in BPM) using other four recently developed algorithms namely SpaMA~\citep{salehizadeh2015novel}, WFPV~\citep{temko2015estimation}, JOSS~\citep{zhang2015photoplethysmography}, and TROIKA~\citep{r10} is also given for comparison. We can see that HSUM gives an improvement over these algorithms. On the other hand, Table~\ref{std_error} shows the standard deviation in error (in BPM) for HR estimates obtained from the HSUM model, HSUM model with 3-point median filtering compared with the HR estimates obtained from the STFT of the PPG signal, and using 3-point median filtering on HR estimates obtained from the STFT of the PPG signal. 
 
The Bland-Altman plot~\citep{r25}for all 12 data-sets is shown in Figure~\ref{fig_ba}. The Bland-Altman plot was used for validating the agreement between the estimated HR values obtained using the HSUM model with the ground-truth HR obtained from the ECG signal~\citep{r25}. The 95 percent limit of agreement (LOA) is [-6.7086, 6.7086] BPM in Bland-Altman plot~\citep{r25}. Figure~\ref{fig_pear} shows the scatter plot of HR estimates with respect to the corresponding ground-truth. The Pearson's coefficient between the HR estimates and the ground-truth is 0.9974. The heart rate estimated using the HSUM model is in agreement with the ground-truth as depicted by the Bland-Altman plot in Figure~\ref{fig_ba}. However, since the estimated HR, as well as ground-truth, are non-normal, Pearson's correlation is not a suitable metric~\citep{r40}. In this paper, we computed Spearman's rank correlation between the HR estimates and the ground-truth as it is applicable for non-normal distributions and is robust to outliers unlike Pearson's correlation~\citep{r39}. The Spearman's rank correlation for all 12 data-sets comes out to be 0.9978 for the HSUM model that shows almost perfect agreement between the HR estimates and the ground-truth.
\section{Conclusions}
We have  developed a harmonic sum-based method for estimation of heart rate (HR) using a PPG signal that had been corrupted by motion-artifacts during running on a treadmill. An auxiliary accelerometer is used to acquire information about the physical movements of the user. The quasi-periodic accelerometer signal is first modeled using a harmonic sum (HSUM) that estimates the fundamental frequency of the acceleration signal over short time-windows. The fundamental frequency of the acceleration signal is then used to model the PPG signal containing information about the heart rate and the motion-artifacts. We extract the heart rate (HR) using a joint HSUM model for the PPG signal. The method is suitable for wearable devices such as PPG wristbands used for real-time HR monitoring. 
\section*{Acknowledgment}
Authors would like to thank anonymous reviewers and editor for helpful comments and valuable suggestion that helped in improving the quality of this paper. 
  
\bibliographystyle{spbasic}
\bibliography{references2}
\end{document}